\documentclass[a4paper,11pt]{article}
\usepackage{pos}

\usepackage[width=.9\textwidth]{subcaption}
\usepackage{graphicx}
\usepackage{siunitx} 

\title{Detection of Small Scale Components in Power Law Spectra}
 \ShortTitle{Detection of Small Scale Components}

\author*[a]{Tim Ruhe}
\author[a]{Wolfgang Rhode}

\affiliation[a]{TU Dortmund University,\\
  Otto Hahn-Straße 4a, 44227 Dortmund, Germany}

\emailAdd{tim.ruhe@tu-dortmund.de}
\emailAdd{wolfgang.rhode@tu-dortmund.de}

\abstract{Spectra in astroparticle physics are commonly approximated by simple power laws. The steeply falling nature of these power laws, however, makes the detection of additional components rather challenging. This holds true especially, if the additional components are small compared to the established ones. Energy spectra of muon neutrinos are an interesting example of such a scenario, where the conventional and astrophysical components to the spectra have been established by the use of different analysis methods, such as likelihood fits or spectral deconvolution. The prompt component, although expected from theoretical models, has not yet been experimentally observed. Furthermore, the extraction of physics parameters is challenged by the large systematic uncertainties, especially at high energies.
This contribution presents a different approach to the analysis of power-law spectra, which is based on functional data analysis. The method itself and its implications are discussed using muon and neutrino energy spectra as an example.}

\FullConference{37$^{\rm{th}}$ International Cosmic Ray Conference (ICRC 2021)\\
		July 12th -- 23rd, 2021\\
		Online -- Berlin, Germany}

\begin{document}
\maketitle

\section{Introduction}
 
\begin{figure}
    \centering
    \includegraphics[width=.7\linewidth]{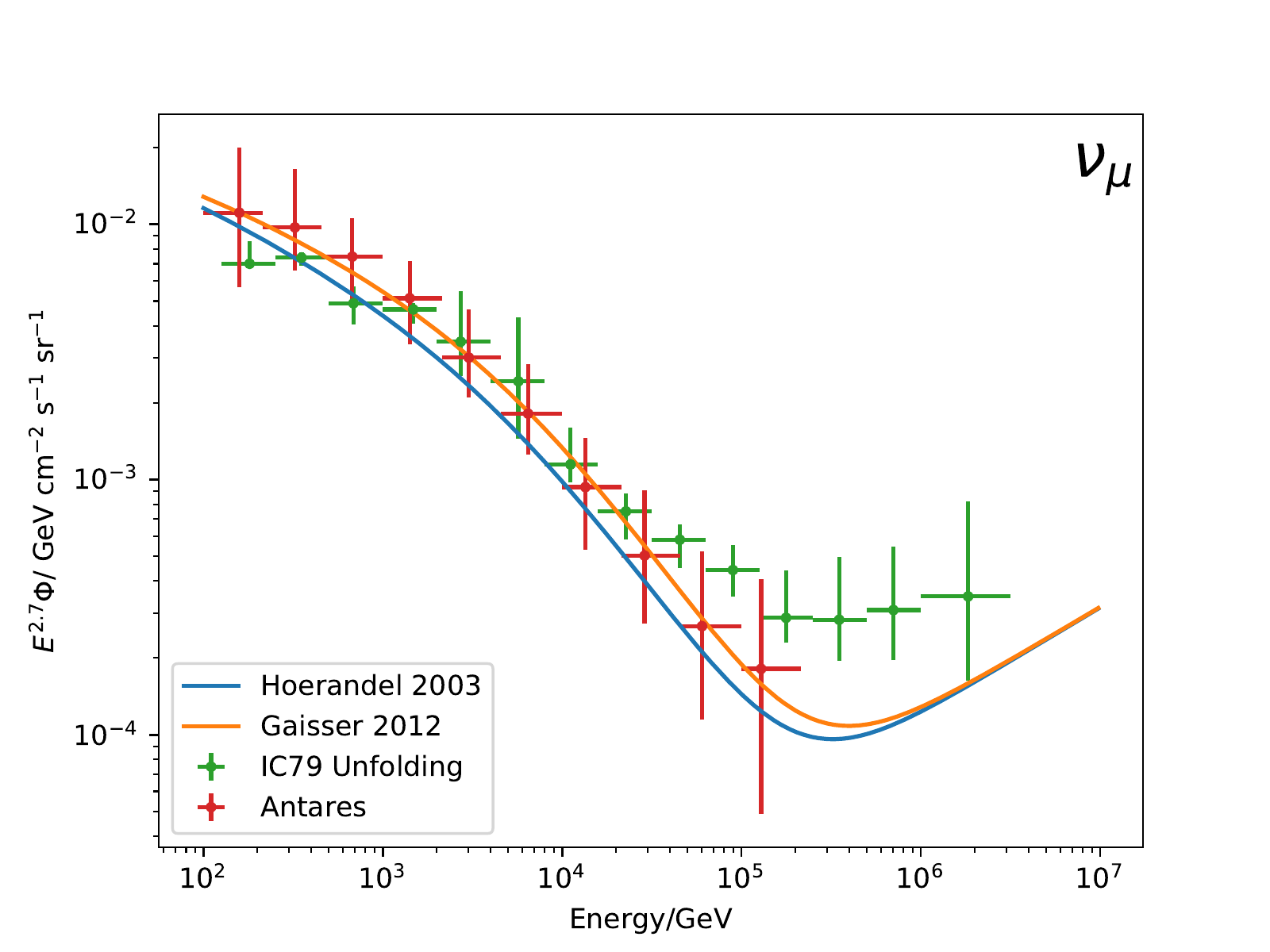}
    \caption{Muon neutrino energy spectra measured by Antares (red) and IceCube (green), compared to theoretical fluxes obtained with MCEq for the cosmic ray models by Gaisser and Hoerandel using SYBILL-2.3c as a hadronic interaction model.}
    \label{fig:atmospheric_spectra}
\end{figure} 
 
The measurement of lepton energy spectra via the use of deconvolution techniques has become a routine task for state-of-the-art neutrino telescopes (cf.~\cite{IC59Atmospheric,IC79Atmospheric,AntaresAtmospheric}). An observed flattening of the spectrum for energies exceeding $10^{5}\,\si{\giga\electronvolt}$ has been attributed to a diffuse flux of high energy neutrinos from astrophysical sources~\cite{IC79Atmospheric}. Flux measurements therefore not only complement other analysis techniques, but also have the capability to confirm results in more model-independent analyses. 

Atmospheric neutrino spectra are expected to consist of three components, the first one being \textit{conventional} atmospheric neutrinos, which originate from the decay of pions and kaons~\cite{gaisser2016,Learned2000,Honda2015}. Due to their relatively long lifetime ($\tau\approx10^{-8}\,\si{\second}$~\cite{PDG}), the parent mesons can lose energy in collision before decaying into neutrinos. Hence, the flux of conventional neutrinos is one power steeper (approximately $\frac{d\Phi}{dE}\propto E^{-3.7}$), compared to the spectrum of the incident cosmic rays. So-called \textit{prompt} neutrinos, which are expected from the decay of charmed mesons, but have not yet been experimentally detected, are the second component~\cite{gaisser2016,Learned2000,berss}. A diffuse flux of neutrinos from astrophysical sources, contributes to the spectrum as a third component(cf.~\cite{Aachen10yrs,HESE7.5}). 

Figure~\ref{fig:atmospheric_spectra} shows muon neutrino energy spectra, obtained with Antares~\cite{AntaresAtmospheric} and IceCube-79~\cite{IC79Atmospheric}. In addition, theoretical predictions obtained with MCEq~\cite{MCEq} for the cosmic ray models by Hoerandel~\cite{Hoerandel2003} and Gaisser~\cite{Gaisser2012} are depicted. For both models, the zenith-averaged muon neutrino fluxes were computed, using SYBILL-2.3c~\cite{Riehn2017} as a hadronic interaction model. The depicted fluxes also include a diffuse flux of astrophysical neutrinos, which is modelled as a power law of the form:
\begin{equation}
\dfrac{d\Phi}{dE}=\Phi_{0,\texttt{astro}} \times \left ( \dfrac{E_{\nu}}{100\,\si{\tera\electronvolt}} \right)^{\gamma_{\texttt{astro}}},
\label{eq:astro_power_law}
\end{equation}
using the best fit parameters reported in~\cite{Aachen10yrs} ($\Phi_{0,\texttt{astro}}=1.44\times 10^{-18}$ and $\gamma_{\texttt{astro}} = -2.28$). From Fig.~\ref{fig:atmospheric_spectra} one finds that although the experimental results agree with the predictions, the discriminative power of the spectra with respect to different cosmic ray models is limited, due to the large uncertainties, especially at high energies. Furthermore, the spectra hardly allow for the extraction of physics parameters, which limits the quantitative comparability to results obtained with different analysis techniques. In addition, a possible contribution of prompt neutrinos cannot be identified.

%Figure~\ref{fig:Minima} depicts the observed minimum of the energy weighted differential muon neutrino flux $E^{m}\frac{\mathop{d\Phi}}{\mathop{dE}}$ as a function of the magnitude $m$. Values obtained using the cosmic ray model by Hoerandel~\cite{Hoerandel} are shown in blue, whereas values obtained using the model by Gaisser~\cite{Gaisser2012} are depicted in green. The model by Fedynitch et al.~\cite{Fedynitch2012} is depicted as well. Again, SYBILL-2.3c~\cite{Riehn2017} was used as the hadronic interaction model. As for Fig.~\ref{fig:AtmosphericSpectra}, the diffuse flux of astrophysical neutrinos was modelled using the values reported in~\cite{Aachen10yrs}. One finds that the minimum of the energy weighted differential muon neutrino flux, decreases with $m$. Furthermore, one observes that the minimum of $E^{m}\frac{\mathop{d\Phi}}{\mathop{dE}}$ differs betweent different cosmic ray models. 

This paper presents a complementary analysis approach, which utilises the positions $E_{\texttt{min}}$ of the minima of energy weighted differential lepton energy spectra $E^m\frac{d\Phi}{dE}$, obtained by neutrino telescopes. In case the position of the minima is known with sufficient precision, the presented method allows for a more accurate discrimination between different cosmic ray models. The method further allows for the detection of additional, but not yet detected components, like the prompt component of the atmospheric muon and neutrino flux.

The paper is organized as follows: Section~\ref{sec:positions_of_mins} investigates the dependency of $E_{\texttt{min}}$ on the weighting factor, as well as on the underlying cosmic ray model. In Sec.~\ref{sec:physics_parameters} the extraction of physics parameters is studied, whereas Sec.~\ref{sec:small_scale_components} discusses possible detection methods for small scale components. Section~\ref{sec:summary} concludes the paper with a discussion and a summary.

\section{Positions of the minima}
\label{sec:positions_of_mins}

\begin{figure}[t]
\centering
\begin{subfigure}[t]{.5\textwidth}
  \centering
  \includegraphics[width=.9\linewidth]{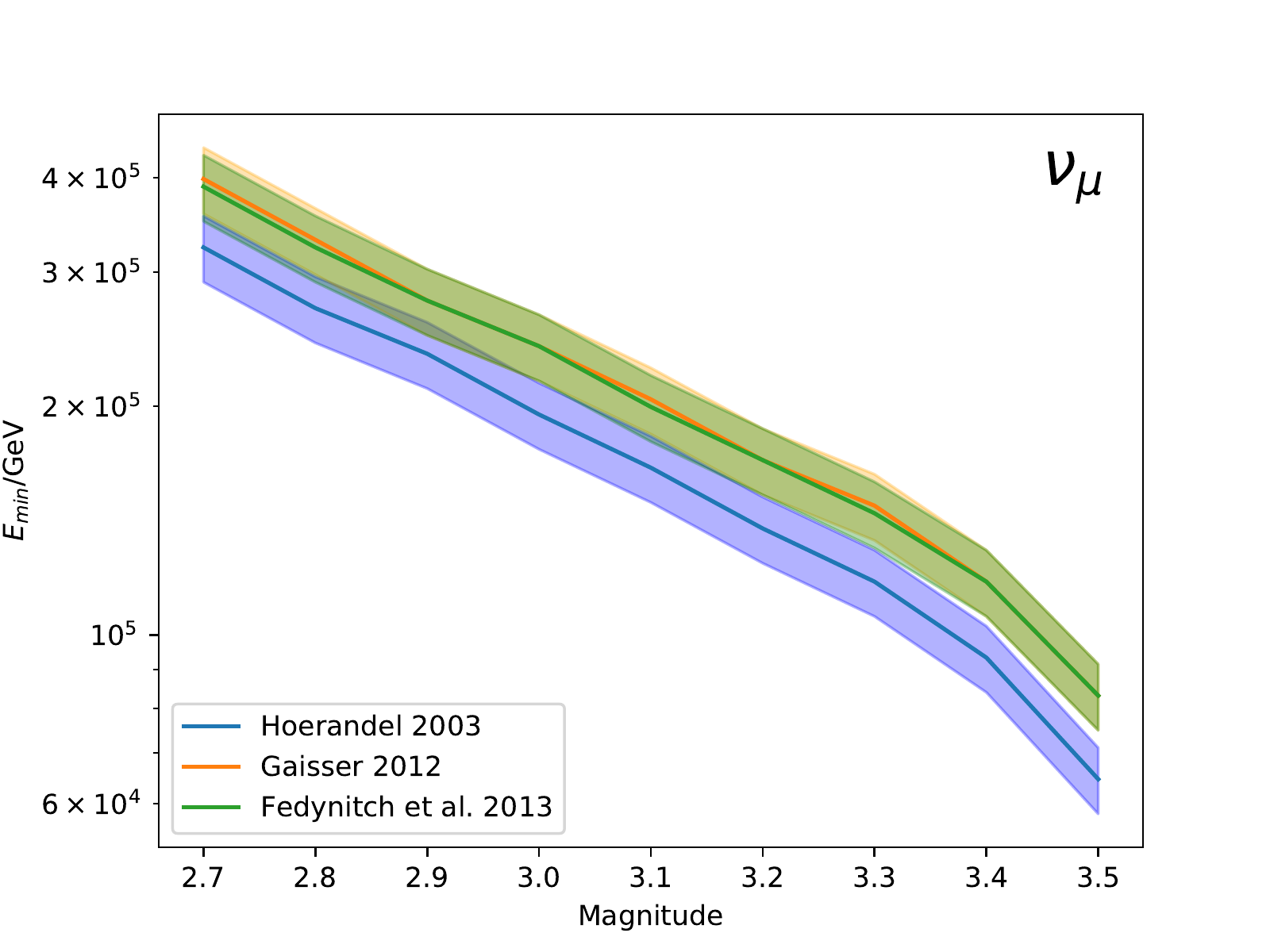}
  \caption{}
  %\caption{Muon Neutrino Energy Spectra as obtmained by IceCube-79 and Antares, compared to theoretical fluxes obtained with MCEq for different models of the cosmic ray flux.}
  \label{fig:Minima1}
\end{subfigure}%
%~
\begin{subfigure}[t]{.5\textwidth}
  \centering
  \includegraphics[width=.9\linewidth]{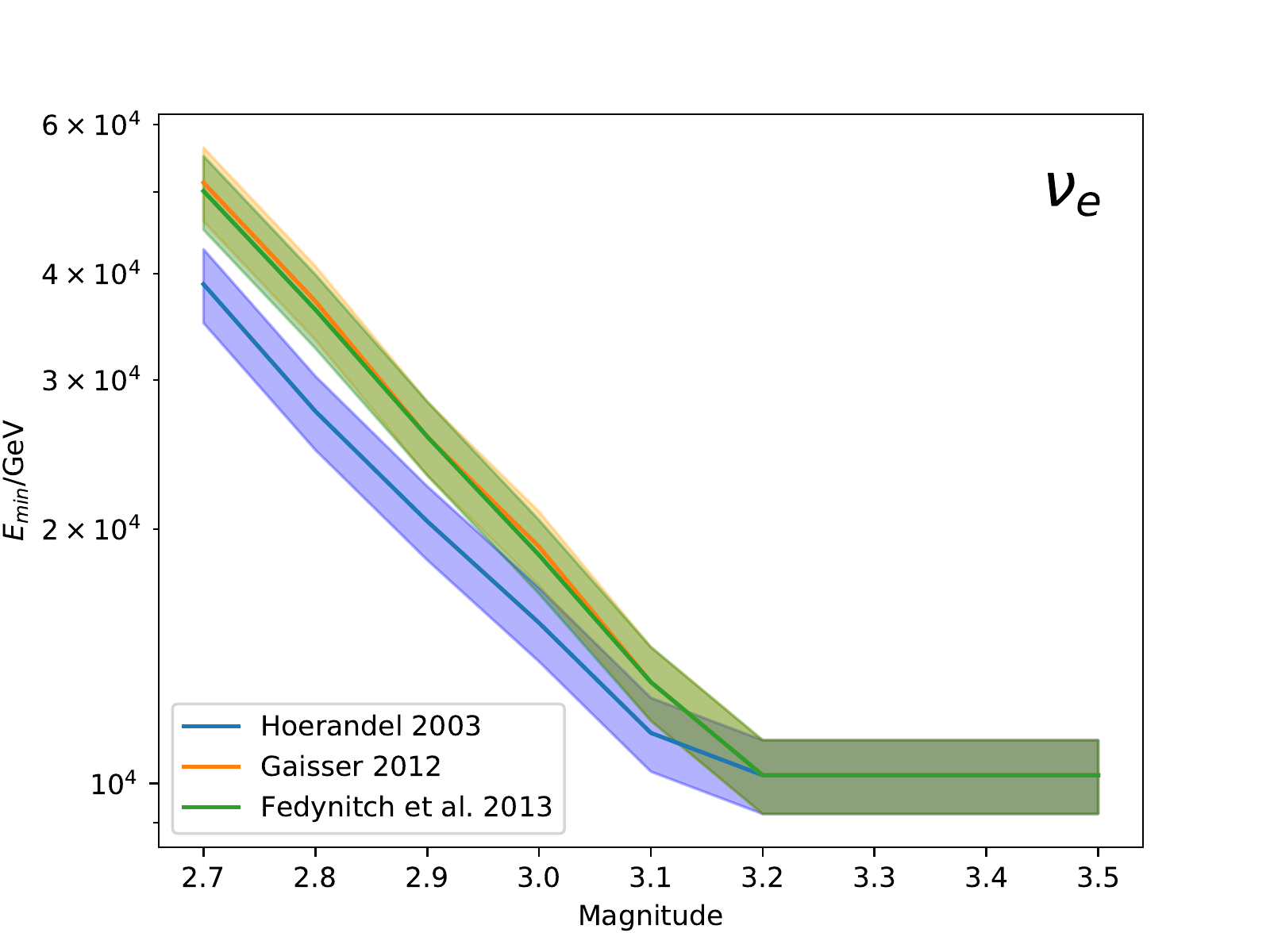}
  \caption{}
  %\caption{Position of the minima of the muon neutrino energy spectrum as a function of the magnitude used for the weighting of the flux.}
  \label{fig:Minima2}
\end{subfigure}
~
\begin{subfigure}[t]{.5\textwidth}
  \centering
  \includegraphics[width=.9\linewidth]{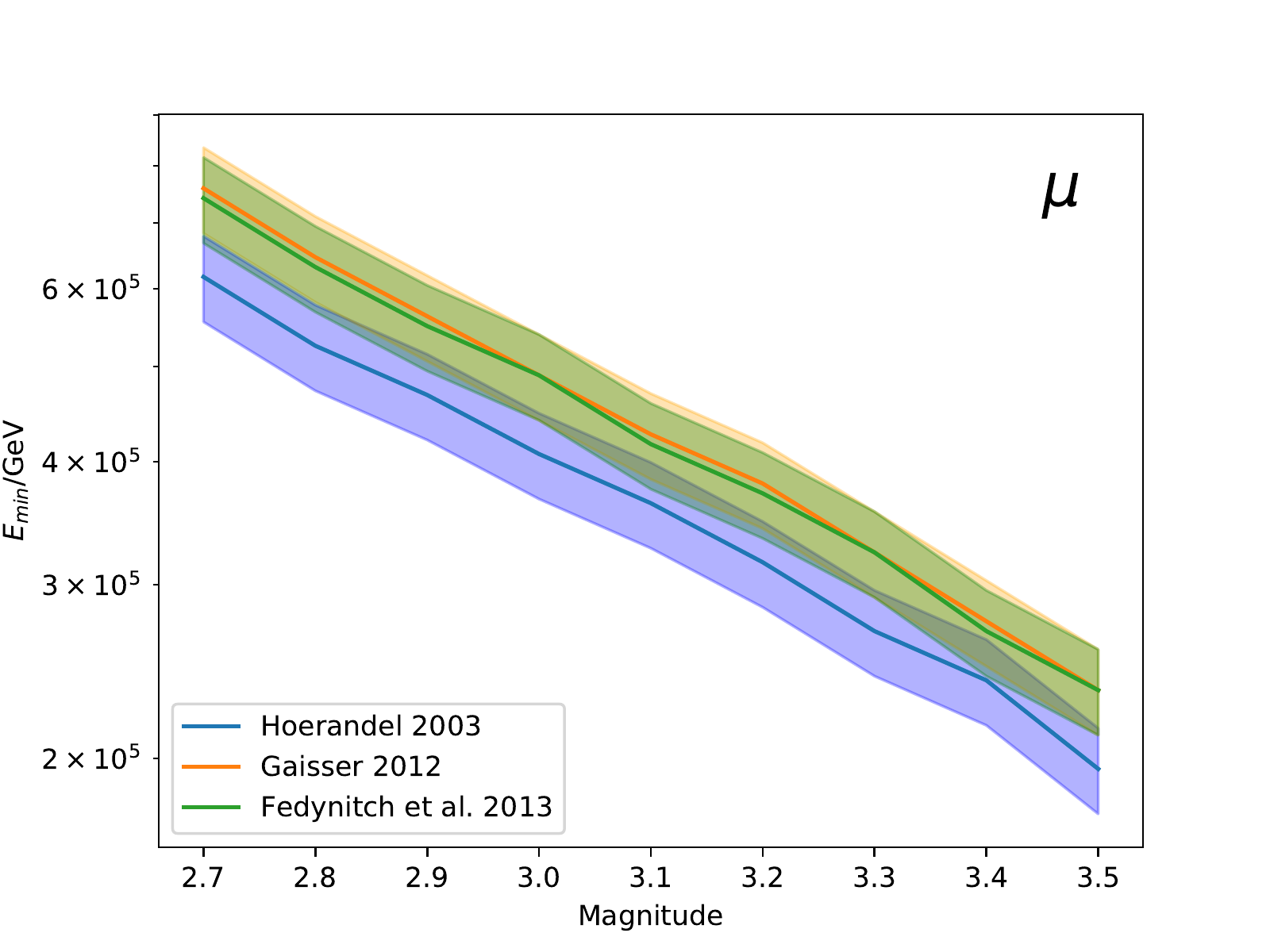}
  \caption{}
  %\caption{Position of the minima of the muon neutrino energy spectrum as a function of the magnitude used for the weighting of the flux.}
  \label{fig:Minima3}
\end{subfigure}
\caption{Position of the minima $E_{\texttt{min}}$ of the energy weighted lepton fluxes $E^m\frac{d\Phi}{dE}$ as a function of $m$, depicted for muon neutrinos (a), electron neutrinos (b) and atmospheric muons (c). }
\label{fig:Minima_vs_m}
\end{figure}
Fig.~\ref{fig:Minima_vs_m} depicts the positions of the mininma $E_{\texttt{min}}$ of energy weighted lepton fluxes $E^m\frac{d\Phi}{dE}$ as a function of $m$ for muon neutrinos (Fig.~\ref{fig:Minima1}), electron neutrinos (Fig.~\ref{fig:Minima2}) and atmospheric muons (Fig.~\ref{fig:Minima3}). The data were obtained using MCEq~\cite{MCEq} for the cosmic ray models by Hoerandel~\cite{Hoerandel2003}, Gaisser~\cite{Gaisser2012} and Fedynitch at al.~\cite{Fedynitch2012}. An uncertainty of 10\% on $E_{\texttt{min}}$ was assumed for all three models. The contribution of astrophysical neutrinos was modelled as a power law (see Eq.~\ref{eq:astro_power_law}), using the best fit parameters reported in~\cite{Aachen10yrs}.

One finds that $E_{\texttt{min}}$ decreases with increasing $m$ for all three cosmic ray models as well as for all three leptons considered. One further finds that $E_{\texttt{min}}$ additionally depends on the underlying cosmic ray model. Evaluating $E_{\texttt{min}}$ as a function of $m$ therefore provides the possibility to distinguish between different cosmic ray models. Such a study would, however, require an increased resolution of the spectra in the region of interest ($10^4$ to $10^6\,\si{\giga\electronvolt}$), which can possibly be provided by the use of machine learning based deconvolution algorithms~\cite{bunse2018}.

\section{Extraction of Physics Parameters}
\label{sec:physics_parameters}

\begin{figure}[t]
\centering
\begin{subfigure}[t]{.5\textwidth}
  \centering
  \includegraphics[width=.9\linewidth]{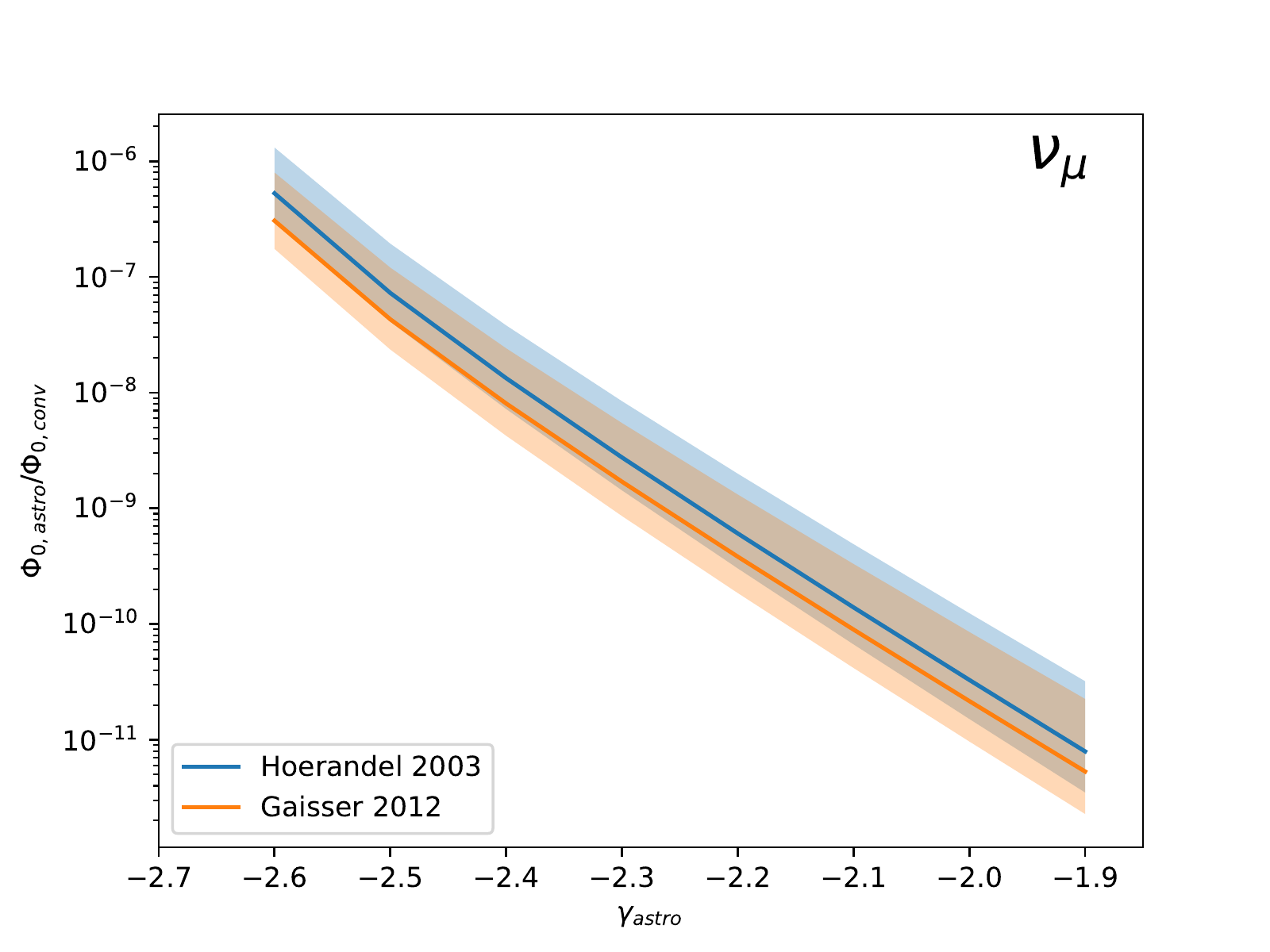}
  \caption{}
  %\caption{Muon Neutrino Energy Spectra as obtmained by IceCube-79 and Antares, compared to theoretical fluxes obtained with MCEq for different models of the cosmic ray flux.}
  \label{fig:Ratio1}
\end{subfigure}%
%~
\begin{subfigure}[t]{.5\textwidth}
  \centering
  \includegraphics[width=.9\linewidth]{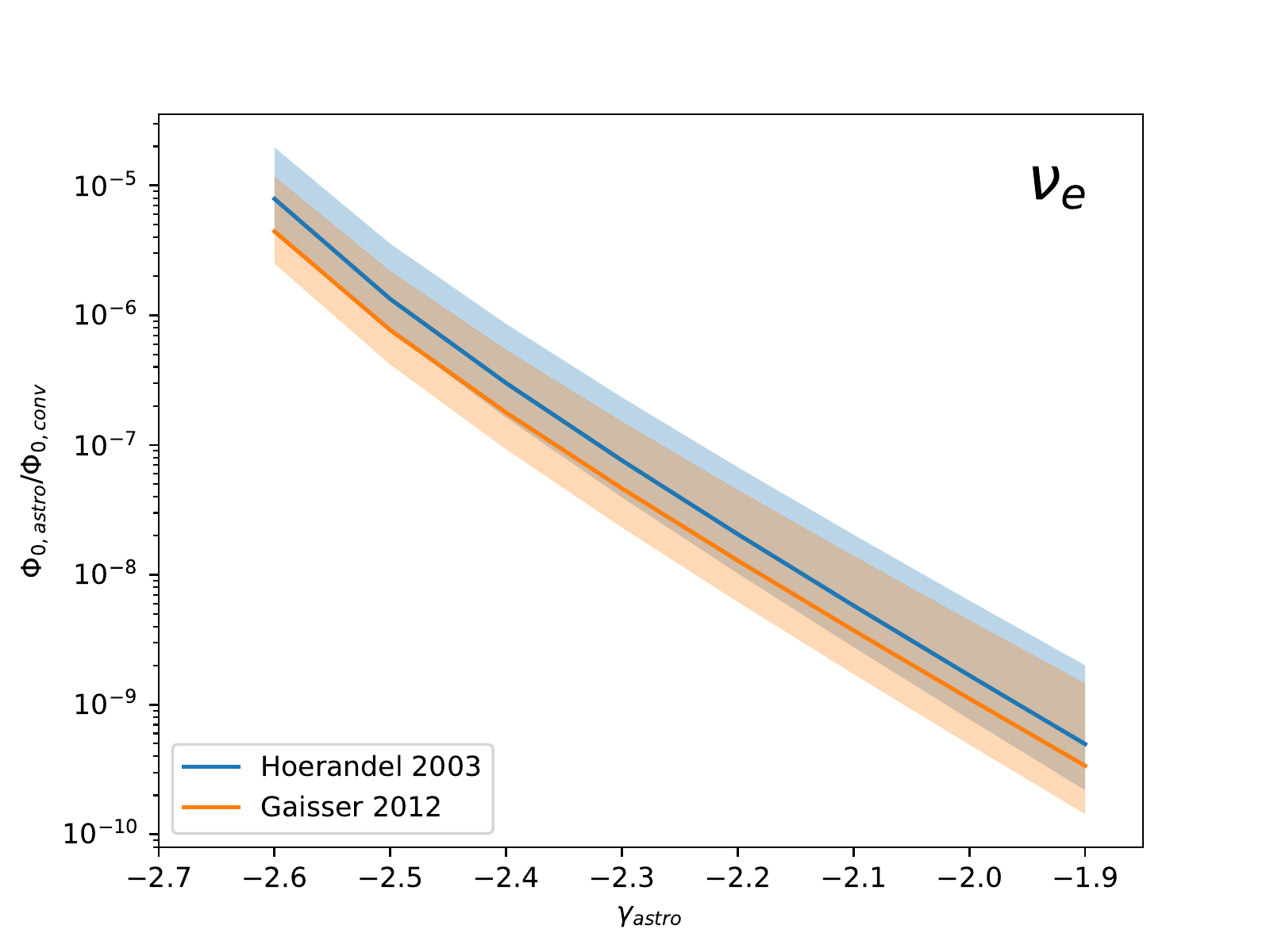}
  \caption{}
  %\caption{Position of the minima of the muon neutrino energy spectrum as a function of the magnitude used for the weighting of the flux.}
  \label{fig:Ratio2}
\end{subfigure}
~
\begin{subfigure}[t]{.5\textwidth}
  \centering
  \includegraphics[width=.9\linewidth]{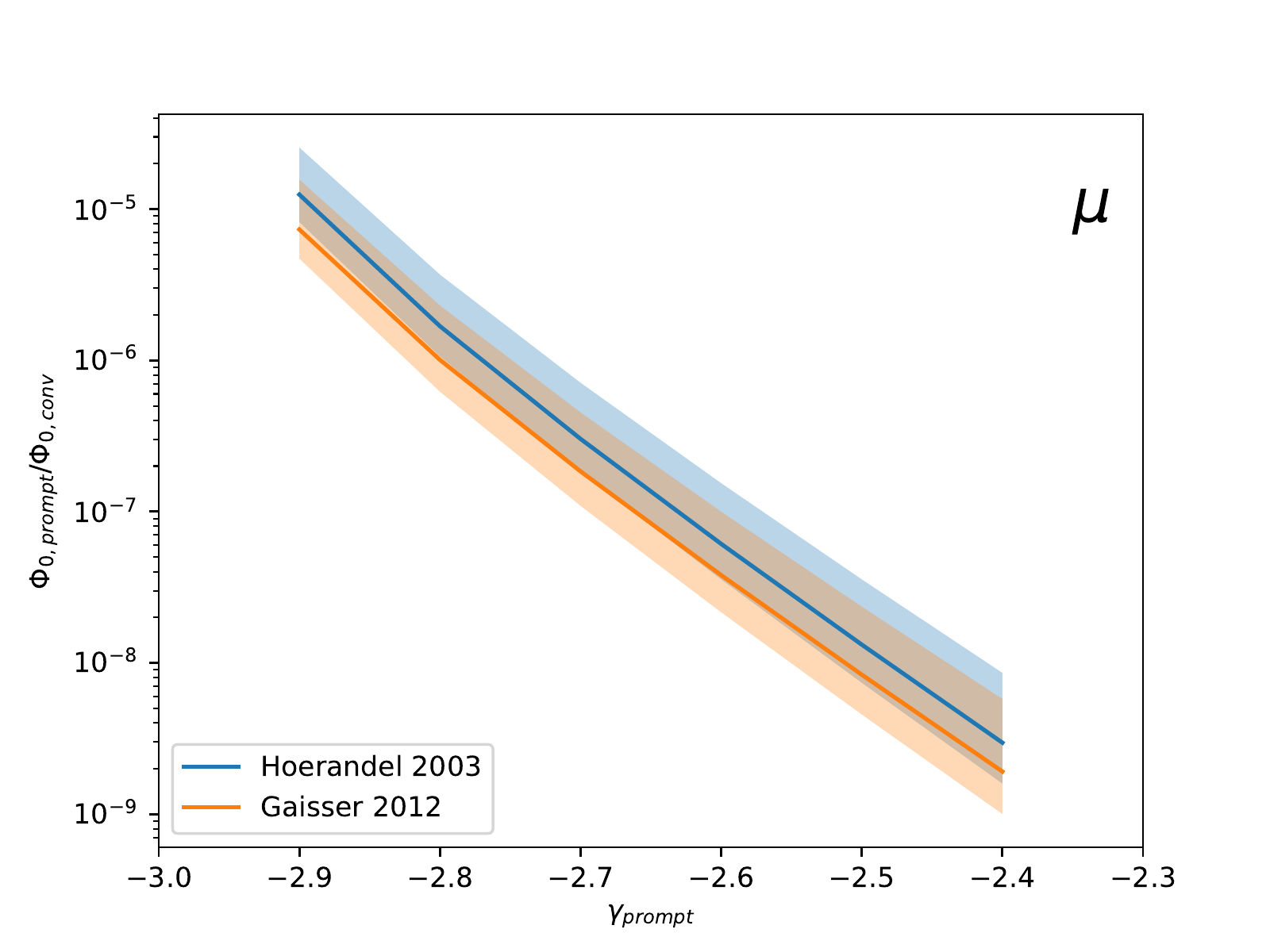}
  \caption{}
  %\caption{Position of the minima of the muon neutrino energy spectrum as a function of the magnitude used for the weighting of the flux.}
  \label{fig:Ratio}
\end{subfigure}
\caption{Ratios of $\Phi_{0,\texttt{astro}}$ (a and b) and $\Phi_{0,\texttt{prompt}}$ (c) to $\Phi_{0,\texttt{conv}}$ as a function of $\gamma_{\texttt{astro}}$ and $\gamma_{\texttt{prompt}}$, respectively.}
\label{fig:Ratios}
\end{figure}
Minima of energy weighted lepton spectra can further be used for the measurement of physics parameters, like the ratio of the normalisations of a component of interest and the conventional component ($\Phi_{0,\texttt{int}}/\Phi_{0,\texttt{conv}}$). The component of interest could be the astrophysical component for electron and muon neutrinos or the prompt component for atmospheric muons. 

Modelling the fluxes as a sum of two power laws (the contribution of the prompt component is neglected for the neutrino case) and differentiating by the energy one obtains:
\begin{equation}
    \dfrac{d\Phi^\prime(E)}{dE} = \Phi_{0,\texttt{conv}}(m + \gamma_{\texttt{conv}})E^{m + \gamma_{\texttt{conv}} - 1} + \Phi_{0,\texttt{int}}(m + \gamma_{\texttt{int}})E^{m + \gamma_{\texttt{int}}- 1},
\end{equation}
where
\begin{equation}
    \Phi^\prime(E) = E^{m}\dfrac{d\Phi}{dE}.
\end{equation}
Demanding
\begin{equation}
    \dfrac{d\Phi^\prime(E) }{dE} \overset{!}{=} 0
\end{equation}
and solving for $\Phi_{0,\texttt{int}}/\Phi_{0,\texttt{conv}}$ yields:
\begin{equation}
\dfrac{\Phi_{0,\texttt{int}}}{\Phi_{0,\texttt{conv}}} = E^{\gamma_{\texttt{conv}} - \gamma_{\texttt{int}}} \dfrac{(\gamma_{\texttt{conv}} + m)}{-(\gamma_{\texttt{int}}+m)}.
\label{eq:2d}
\end{equation}
$\Phi_{0,\texttt{int}}/\Phi_{0,\texttt{conv}}$ thus becomes a function of $E_{\texttt{min}}$ and $\gamma_{\texttt{int}}$. Accordingly, Fig.~\ref{fig:Ratios} depicts $\Phi_{0,\texttt{int}}/\Phi_{0,\texttt{conv}}$ as a function of  $\gamma_{\texttt{astro}}$ for $\nu_{\mu}$ and $\nu_e$ and as function of  $\gamma_{\texttt{prompt}}$ for atmospheric muons. Again, the plots were obtained using MCEq~\cite{MCEq} for the cosmic ray models by Hoerandel and Gaisser, utilising SYBILL-2.3c as a hadronic interaction model. The contribution of astrophysical neutrinos was modelled as a power law, using the best fit parameters reported in~\cite{Aachen10yrs}. An uncertainty of 50\% was assumed on $E_{\texttt{min}}$, which approximately corresponds to the energy resolution reported in~\cite{IC79Atmospheric}. $\Phi_{0,\texttt{conv}}$ and $\gamma_{\texttt{conv}}$ were obtained by a power law fit to the simulated fluxes. The fit was restricted to the energies between $10^4$ and $10^6\,\si{\giga\electronvolt}$.

\section{Detection of Small Scale Components}
\label{sec:small_scale_components}

\begin{figure}[t]
\centering
\begin{subfigure}[t]{.5\textwidth}
  \centering
  \includegraphics[width=.9\linewidth]{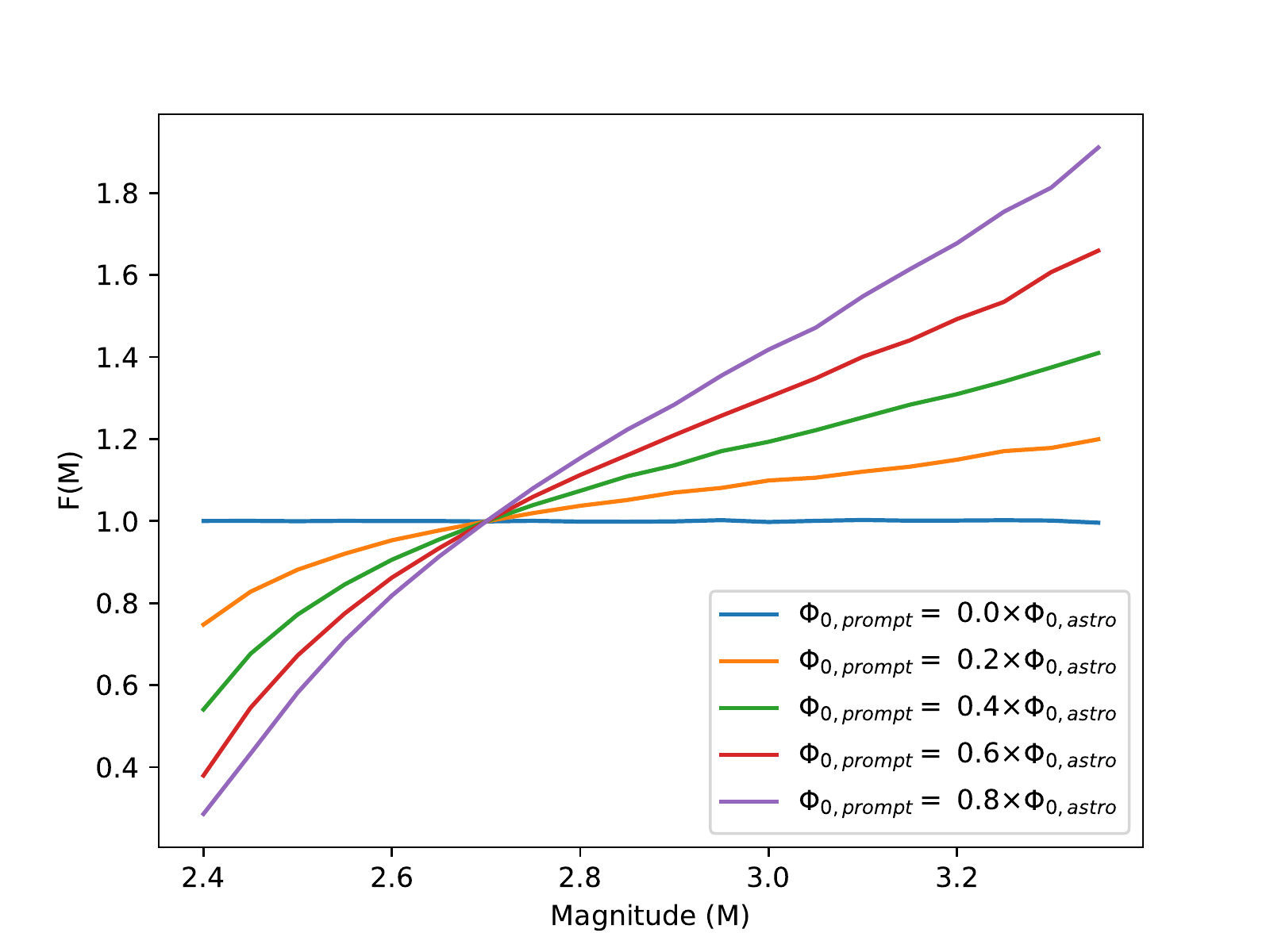}
  \caption{}
  %\caption{Muon Neutrino Energy Spectra as obtmained by IceCube-79 and Antares, compared to theoretical fluxes obtained with MCEq for different models of the cosmic ray flux.}
  \label{fig:F_vs_m}
\end{subfigure}%
%~
\begin{subfigure}[t]{.5\textwidth}
  \centering
  \includegraphics[width=.9\linewidth]{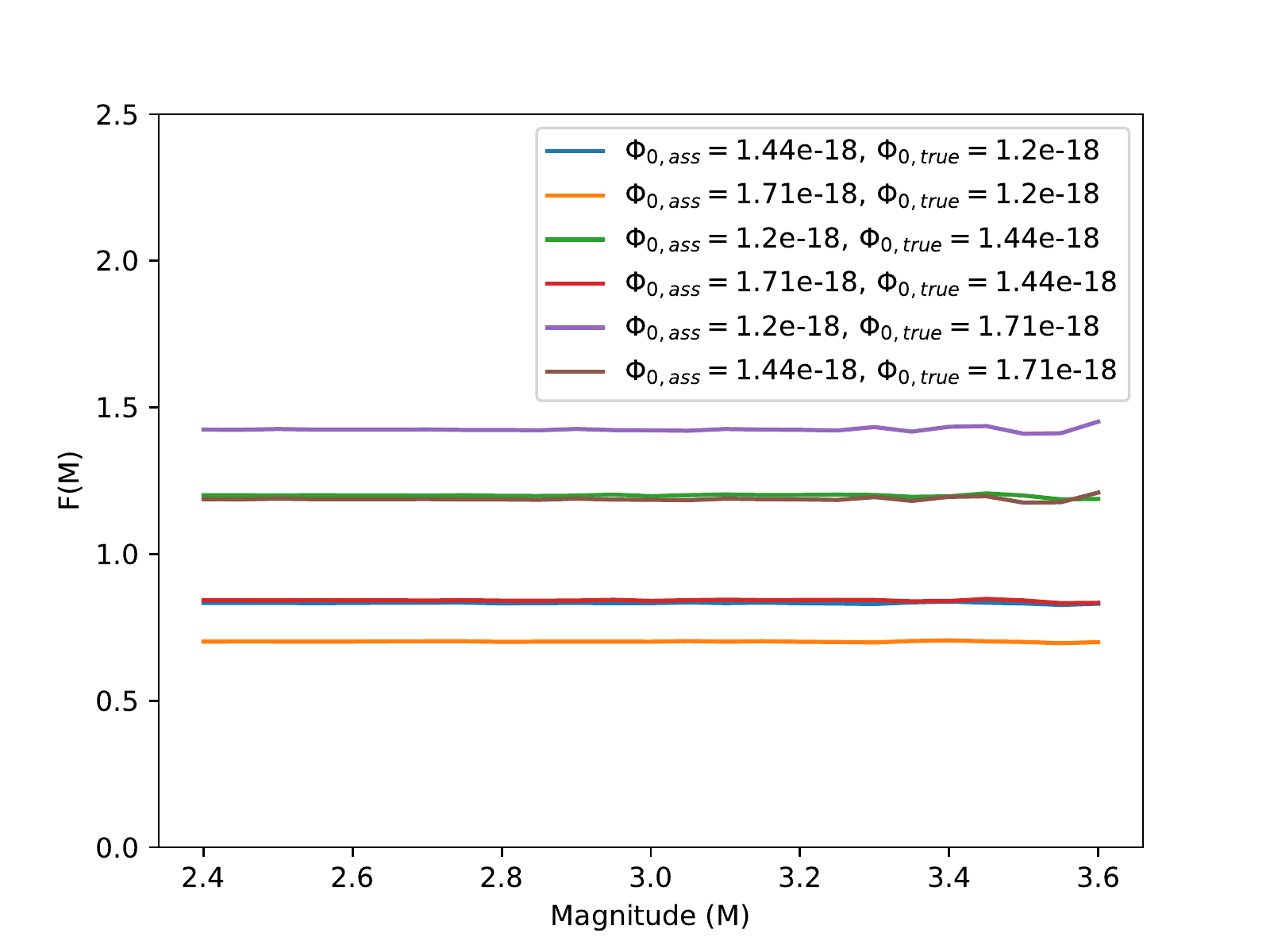}
  \caption{}
  %\caption{Position of the minima of the muon neutrino energy spectrum as a function of the magnitude used for the weighting of the flux.}
  \label{fig:norm_dep}
\end{subfigure}
~
\begin{subfigure}[t]{.5\textwidth}
  \centering
  \includegraphics[width=.9\linewidth]{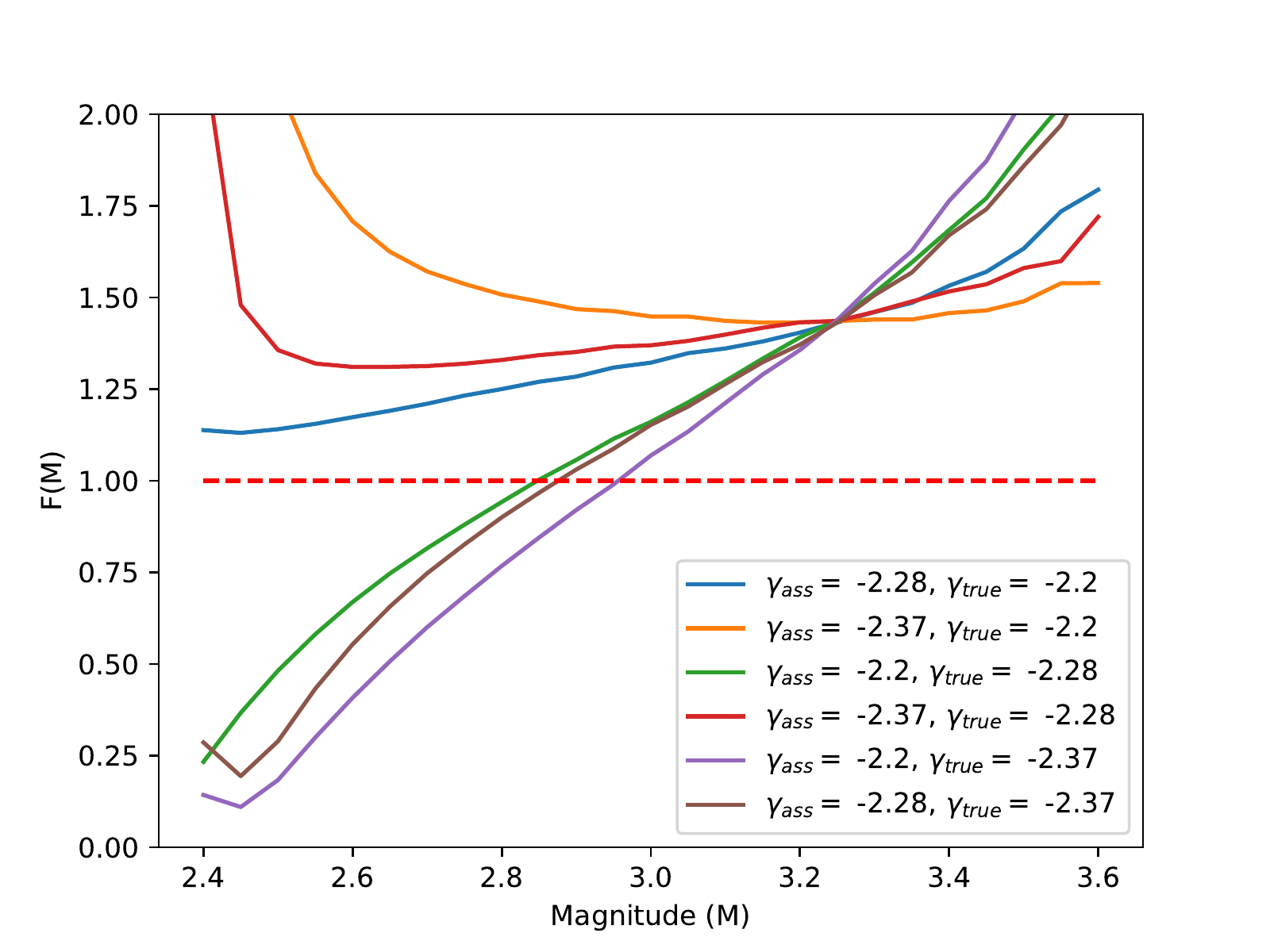}
  \caption{}
  %\caption{}
  \label{fig:gamma_dep}
\end{subfigure}
\caption{The derived quantity $F$ as a function of the magnitude $m$, showing the impact of the prompt component (a), as well as the impact of uncertainties in the normalisation (b) and the spectral index (c) of the astrophysical component.}
\label{fig:small_scale}
\end{figure}
Assuming a contribution of only two components (conventional and astrophysical) to the lepton fluxes and using the same approach as in Sec.~\ref{sec:physics_parameters}, one obtains:
\begin{equation}
 \Phi_{0,\texttt{conv}}(m + \gamma_{\texttt{conv}})E^{w+\gamma_{\texttt{conv}} - 1} = \Phi_{0,\texttt{astro}}(m + \gamma_{\texttt{astro}})E^{w + \gamma_{\texttt{astro}}- 1},
\end{equation}
which can be re-written as:
\begin{equation}
    F(w) := \dfrac{- \Phi_{0,\texttt{conv}}(m + \gamma_{\texttt{conv}})}{\Phi_{0,\texttt{astro}}(m + \gamma_{\texttt{astro}})} E^{\gamma_{\texttt{astro}} - \gamma_{\texttt{conv}}} = 1
    \label{eq:F}
\end{equation}
Considering Eq.~\ref{eq:F} one finds that the presence of an additional component will cause $F(m)$ to deviate from one. The size of the deviation, however, depends on the normalisation of the additional component. This is illustrated in Fig.~\ref{fig:F_vs_m}, which depicts $F(m)$ as a function of $m$. To obtain this figure, all components were modelled as power laws and the parameters of the astrophysical neutrino flux were chosen according to the values reported in~\cite{Aachen10yrs}. For the prompt component a spectral index of $\gamma_{\texttt{prompt}} = -2.7$ was used and $\Phi_{0,\texttt{prompt}}$ was expressed in units of $\Phi_{0,\texttt{astro}}$. The conventional component was modelled as $\frac{d\Phi_{\texttt{conv}}}{dE}=\Phi_{0,\texttt{conv}} \times \left ( \frac{E_{\nu}}{100\,\si{\tera\electronvolt}} \right)^{\gamma_{\texttt{conv}}}$, with $\Phi_{0,\texttt{conv}} = 1.\times 10^{-18} $ and $\gamma_{conv}= -3.7$. One finds that a horizontal line is observed for a vanishing prompt component, whereas increasing deviations from said line were observed for increasing values of $\Phi_{0,\texttt{prompt}}$.

As the parameters of the astrophysical flux are subject to systematic uncertainties, the impact of these uncertainties was exemplarily studied for $\Phi_{0,\texttt{astro}}$, while assuming negligible uncertainties on $\gamma_{\texttt{astro}}$. The outcome of these studies is depicted in Fig.~\ref{fig:norm_dep}. The depicted graphs correspond to various combinations of assumed and true normalisations. Looking at Fig.~\ref{fig:norm_dep} one finds that differences in the true and assumed normalisations only affect the bias of $F$. The shape of the distribution (a horizontal line) remains unaffected. 

Furthermore, the impact of uncertainties in $\gamma_{\texttt{astro}}$ was investigated, while assuming negligible uncertainties on the normalisation. The outcome of these investigations is shown in Fig.~\ref{fig:gamma_dep}. Considering Fig.~\ref{fig:gamma_dep} one finds that in addition to the bias of $F$, the shape of the graph is affected by the uncertainties. Compatring Figs.~\ref{fig:gamma_dep} and~\ref{fig:F_vs_m} one also finds that the shapes of graphs with a difference between the true and the assumed spectral index might not be distinguishable from graphs with a non-vanishing prompt component, in case $\gamma_{\texttt{astro}}$ is assumed to be smaller than its true value. The graphs may, however, be well distinguishable in the opposite case. 

\section{Summary and Discussion}
\label{sec:summary}

In these proceedings we presented an analysis approach based on functional data analysis and the positions $E_{\texttt{min}}$ of the minima of energy weighted lepton spectra $E^m\frac{d\Phi}{dE}$. The presented approach leverages the fact that $E_{\texttt{min}}$ depends on $m$, as well as on the underlying cosmic ray model, which allows for a more accurate discrimination between cosmic ray models, compared to the spectra alone. 

We further showed, how $E_{\texttt{min}}$ can be used for the extraction of physics quantities, like the ratio $\Phi_{0,\texttt{int}}/\Phi_{0,\texttt{conv}}$ of the normalisations of a component of interest $\Phi_{\texttt{int}}$ and the conventional component $\Phi_{\texttt{conv}}$. We showed that $\Phi_{0,\texttt{int}}/\Phi_{0,\texttt{conv}}$ becomes a function of the spectral index $\gamma_{\texttt{int}}$ of said component. This dependency allows for the extraction of physics parameters, which can otherwise not be accessed from the spectra. The extracted quantities further provide the possibility for a quantitative comparison between spectral measurements and other analyses, e.g. likelihood fits, which can otherwise only be compared on a qualitative level.

As a third study we showed, that the positions of the minima can also be used for the detection of additional components via a quantity $F(m)$ (see Eq.~\ref{eq:F}). $F(m)$ equals one, for all $m$ in case no more than two components constitute the overall lepton flux, when both components are modelled as power laws. $F(m)$ is significantly altered in case additional components contribute to the spectrum. Although, the shape of $F$ is robust against uncertainties in the normalisations of the two assumed fluxes, it is affected by the uncertainties in the spectral indices to a certain extent. Upcomig MCEq-based studies on $F$, are expected to provide deeper insight into the applicability of this method in physics measurements, especially on the achievable accuracy given the current uncertainties on the spectral index and the normalisation of the astrophysical neutrino flux. 

In summary, we showed two studies, which indicate that considering the positions of the minima of energy-weighted lepton spectra will be beneficial for the extraction of physics parameters, as well as for the discrimination between models. We further discussed an analsysis method, which will allow for the detection of small scale components in power-law spectra, which would remain undetected in case the minima of the spectra for different weighting factors are not considered in addition to the spectra themselves.

\bibliographystyle{JHEP}       % APS-like style for physics
\bibliography{references}   % name your BibTeX data base

%\begin{thebibliography}{99}
%\bibitem{...}
%....

%\end{thebibliography}

%% Full authors list (ONLY FOR COLLABORATIONS)
%\clearpage
%\section*{Full Authors List: \Coll\ Collaboration}
%
%\noindent \textbf{Note comment afterwards:} Collaborations have the possibility to provide an authors list in xml format which will be used while generating the DOI entries making the full authors list searchable in databases like Inspire HEP. For instructions please go to icrc2021.desy.de/proceedings or contact us under icrc2021proc@desy.de.\\
%
%\scriptsize
%\noindent
%first.author$^1$, 
%second.author$^2$, 
%third.author$^3$ % .... more names
%and 
%last.author$^{n}$ \\
%
%\noindent
%$^1$first.affiliation.
%$^2$second.affiliation. % .... more affiliation
%$^{m}$last.affiliation.

\end{document}